\documentclass{PoS}

\title{Chiral extrapolations and strangeness in the baryon ground states}

\ShortTitle{Chiral extrapolations of baryon masses}

\author{\speaker{Matthias F.M. Lutz}\\
        GSI Helmholtzzentrum f\"ur Schwerionenforschung GmbH\\ Planckstr. 1, 64291 Darmstadt, Germany \\
        E-mail: \email{m.lutz@gsi.de}}

\author{Alexander Semke\\
         GSI Helmholtzzentrum f\"ur Schwerionenforschung GmbH\\ Planckstr. 1, 64291 Darmstadt, Germany \\
        E-mail: \email{a.semke@gsi.de}}

\abstract{We review the quark-mass dependence of the baryon octet and decuplet masses as obtained from 
recent lattice simulations of the BMW, PACS-CS, LHPC, HSC  and QCDSF-UKQCD groups. Our discussion relies on
the relativistic chiral Lagrangian and large-$N_c$ sum rule estimates of the counter terms relevant for the
baryon masses at N$^3$LO. A partial summation is implied by the use of physical baryon and meson masses in the
one-loop contributions to the baryon self energies. In our analysis the physical masses are reproduced 
exactly by means of a suitable set of linear constraints. A quantitative and simultaneous description of all
lattice results is achieved in terms of a six parameter fit, where the symmetry conserving counter term that are
relevant at N$^3$LO are not yet being used. For pion masses larger than 300 MeV there appears to be an
approximate linear pion-mass dependence of all octet and decuplet baryon masses. We discuss the pion- and
strangeness sigma terms of the baryon octet states.}

\FullConference{The 7th International Workshop on Chiral Dynamics,\\
		August 6 -10, 2012\\
		Jefferson Lab, Newport News, Virginia, USA}

\begin{document}

\section{Introduction}

The purpose of this talk is to review the chiral extrapolation and strangeness content of the baryon octet and
decuplet ground states. We discuss the recent lattice data on the pion-mass dependence
of the baryon masses \cite{LHPC2008,PACS-CS2008,HSC2008,BMW2008,Alexandrou:2009qu,Durr:2011mp,WalkerLoud:2011ab,Bietenholz:2011qq}.
Given a systematic analysis based on the chiral Lagrangian, the strange-quark mass
dependence of the baryon masses can be calculated. Such results scrutinize the consistency of the chiral extrapolation approach
and lattice simulations of the baryon masses. Variations of the baryon masses along suitable pathes in the pion-kaon mass plane
are important to improve the determination of the low-energy parameters of QCD as encoded into the chiral Lagrangian.

Our work relies on the relativistic chiral Lagrangian with baryon octet and decuplet fields where effects at
N$^3$LO (next-to-next-to-next-leading order) are considered systematically. The details of the approach are published
in  \cite{Semke2005,Semke2007,Semke2012, SemkePLB2012, LutzRC2012}. The chiral extrapolation of baryon masses with strangeness 
content is critically discussed in the literature 
\cite{Jenkins1991,Frink2006,Semke2007,PACS-CS2008,MartinCamalich:2010fp,PhysRevD.81.014503,Geng:2011wq,MartinCamalich:2010zz,Semke2012,WalkerLoud:2011ab}. 
The convergence properties of a strict chiral expansion for the baryon masses with three light flavors are very poor, if existing at all for a physical 
strange quark mass. Different strategies how to perform partial summations or phenomenological adaptations are investigated by several 
groups \cite{Semke2005,Semke2007,Frink2006,PhysRevD.81.014503,MartinCamalich:2010fp}.
A straight forward application of chiral perturbation theory to recent QCD lattice simulations appears futile 
(see e.g.  \cite{LHPC2008,PACS-CS2008,MartinCamalich:2010fp,WalkerLoud:2011ab}).
A crucial element of our scheme is the use of physical baryon masses in the one-loop contribution to the baryon self energies.
Furthermore, the low-energy constants required at N$^3$LO are estimated by sum rules that follow from QCD in the limit of a large number of
colors ($N_c$) \cite{LutzSemke2010,Semke2012}.
Since we obtain a simultaneous and quantitative description of the lattice data of the BMW, PACS-CS, LHPC, HSC and QCDSF-UKQCD groups, we 
find it justified to present detailed results on the strange-quark mass dependence of all members of the baryon octet and decuplet states. 
In particular, we confront our parameter set against recent analyses of the BMW and QCDSF-UKQCD groups on the pion- and strangeness-sigma 
terms of the baryon octet states \cite{Durr:2011mp,Horsley:2011wr,Bali:2011ks}.

\section{Chiral extrapolation of baryon masses}

We consider the chiral extrapolation of the baryon masses to unphysical quark masses. Assuming exact isospin symmetry, the hadron
masses are functions of  $m_u=m_d\equiv m_q$ and $m_s$. The dependence on the light quark masses may be traded against a dependence
on the pion and kaon masses. The 'physical' strange quark mass is determined such that at the physical pion mass the empirical kaon mass is reproduced.
Our approach is detailed in  \cite{Semke2005,Semke2007,Semke2012}. In particular, we assume a quark-mass
dependence of the pion and kaon masses as predicted by $\chi$PT at the next-to-leading order with parameters as recalled in  \cite{Semke2012}.
The baryon masses are computed at N$^3$LO where we use physical baryon and meson masses in the one-loop contributions to the baryon self
energies and assume systematically large-$N_c$ sum rules for the parameter set.

Initially we adjusted the parameter set to the physical masses of the baryon octet
and decuplet states and to the results for the pion-mass dependence of the nucleon and omega masses as predicted
by the BMW group \cite{BMW2008}. Using the isospin average of the empirical baryon masses we derived a system of linear equations
which was used to express eight low-energy constants in terms of the remaining parameters. This provided a significant
simplification of the parameter determination and allowed us to analyze the set of non-linear and coupled equations in great depth \cite{Semke2012}.
An accurate reproduction of the physical baryon masses and lattice data was achieved.
All parameters, except of the symmetry preserving counter terms relevant at N$^3$LO, were considered.
Given the large-$N_c$ sum rules of  \cite{LutzSemke2010}, there are 5 independent parameters only, which are  all set to zero so far.
The latter have a rather minor effect on the baryon masses and can be determined only with very precise lattice data. That initial parameter set
was further slightly adjusted in \cite{SemkePLB2012} to achieve consistency with recent lattice results of the LHPC and PACS-CS 
groups \cite{WalkerLoud:2011ab,PACS-CS2008}. Insisting on a simultaneous description of the BMW and PACS-CS data we unambiguously recover the 
unfitted results of the HSC and QCDSF-UKQCD groups \cite{HSC2008,Bietenholz:2011qq} with an amazing accuracy \cite{LutzRC2012}. We also explored 
the role of the axial coupling constants $F$ and $D$, which we found to be accurately determined from a simultaneous fit of the lattice data \cite{LutzRC2012}.

\begin{figure}
\includegraphics[width=7.4cm,clip=true]{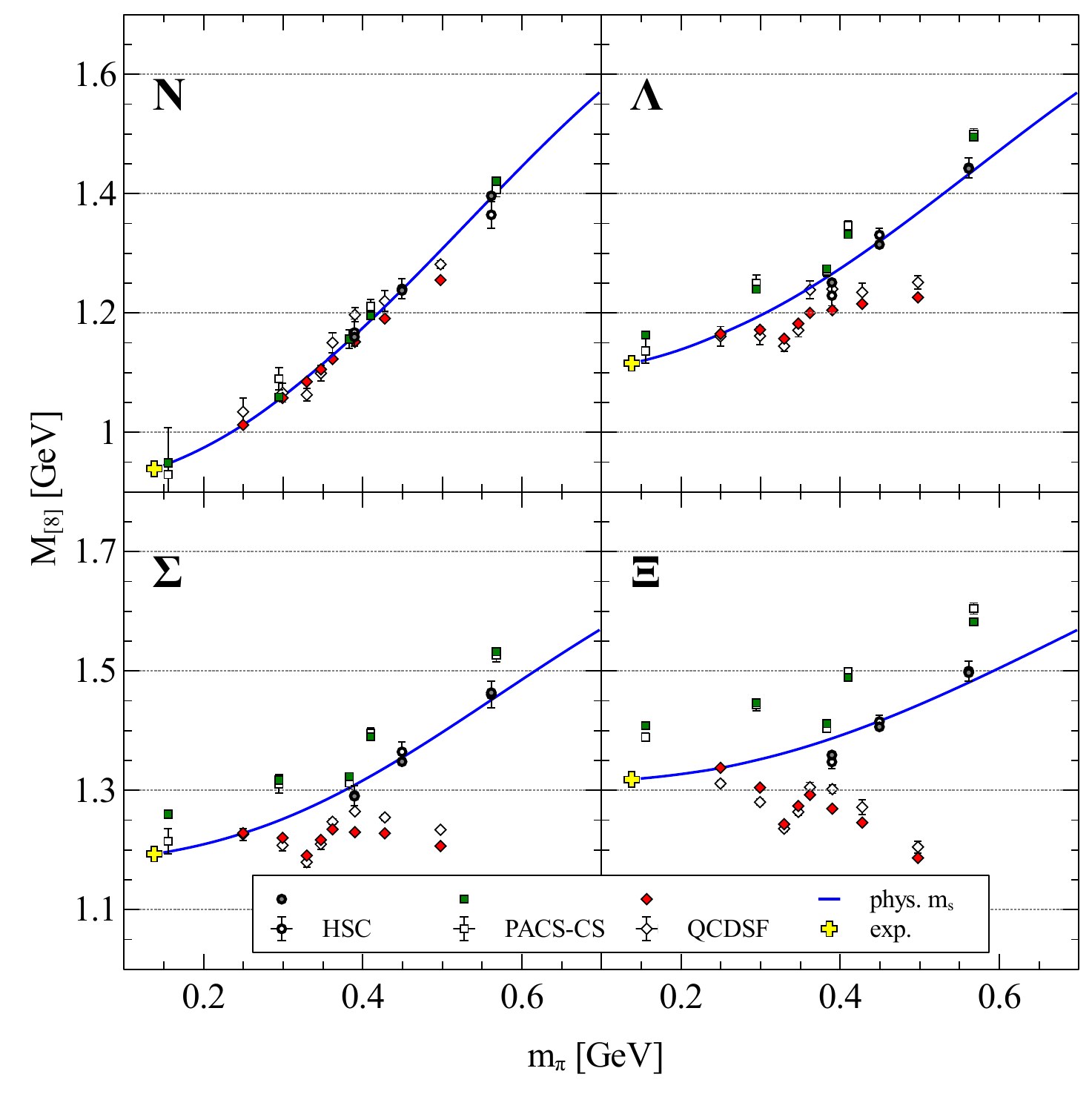} \includegraphics[width=7.4cm,clip=true]{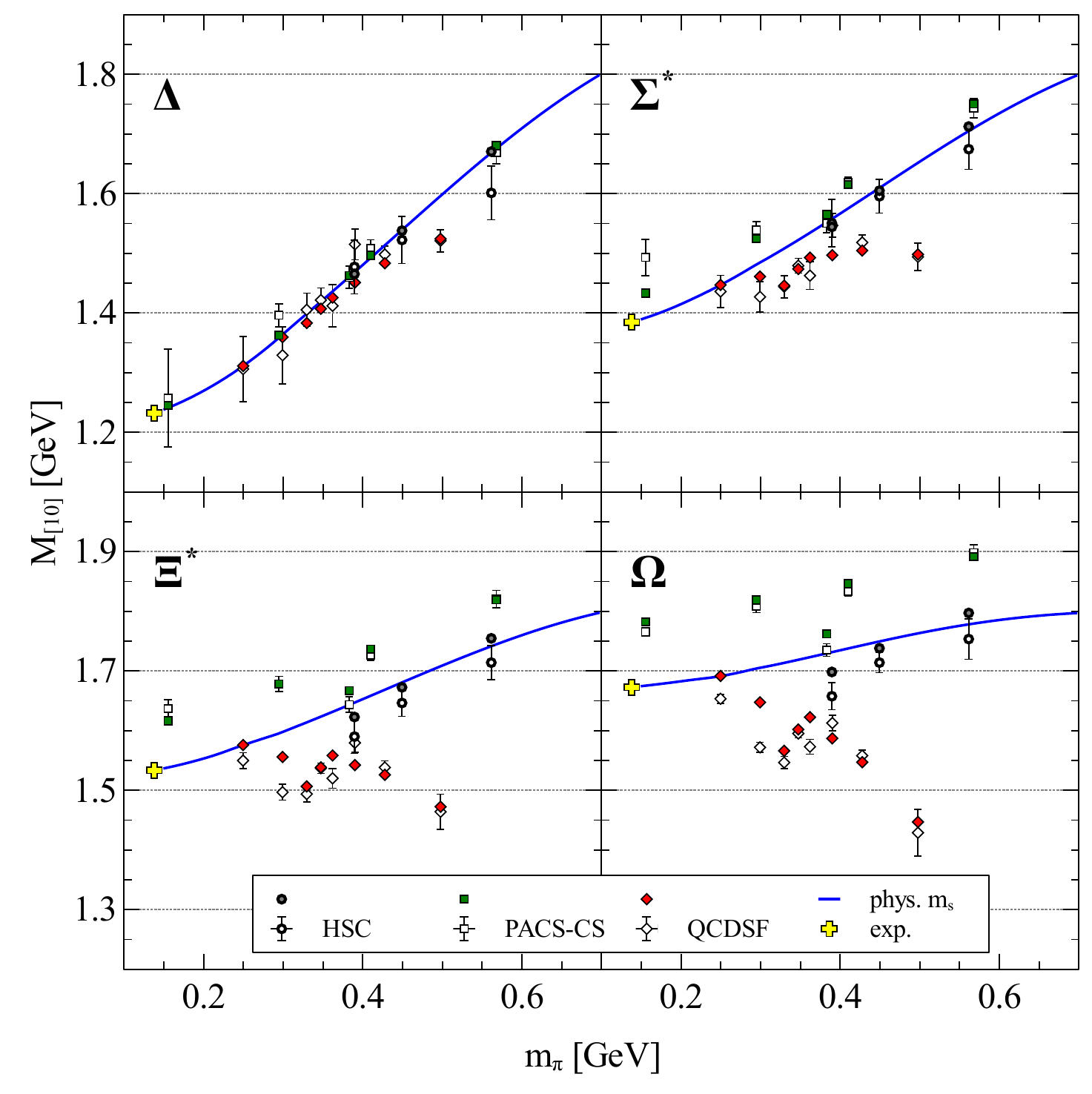}
\caption{Baryon masses as a function of the pion mass as explained in the text. }
\label{figure:baryon-masses}
\end{figure}

In Fig. \ref{figure:baryon-masses} we show the baryon masses for our favored parameter set 3 as a function of the linear pion
mass \cite{SemkePLB2012}. It is a striking phenomena that for pion masses larger than 300 MeV there appears to be an
approximate linear dependence for all baryon masses. Note however, that given our approach the linear dependence is significantly and 
systematically altered at smaller pion masses. Such a behavior was discussed in the talk of Walker-Loud at this conference 
based on various lattice data for the nucleon mass. Our results are confronted against the lattice data 
from the HSC groups (open circles), which provide data for an almost physical strange quark mass. In order to provide a 
quantitative comparison, we compute the baryon masses for the pion and kaon masses as predicted by the HSC group. 
The grey filled circles show our results. The distance of the filled circles 
to the solid lines measures the importance of taking the precise physical strange quark 
mass in the computation of the baryon masses. In Fig. \ref{figure:baryon-masses} we also confront our approach  
with the simulation data of the PACS-CS (open squares) and QCDSF-UKQCD (open diamonds) groups. 
Since the later groups use strange quark masses that are significantly off the physical value, the data points are typically quite 
distant from the solid line. Again our results, green squares and red diamonds, are quite close to their corresponding lattice points, open squares 
and diamonds. Note that all lattice points are shown with statistical errors only, where we assume the central values of the 
lattice spacing as provides by the various lattice groups.

\section{Pion and strangeness baryon sigma terms}

The pion- and strangeness baryon sigma terms play an important role in various physical systems.
For instance, the pion-nucleon sigma term is of greatest relevance in the determination of the
density dependence of the quark condensate at low baryon densities and therefore
provides a first estimate of the critical baryon density at which chiral symmetry may be restored
(see e.g.  \cite{Lutz:1999vc}).  Similarly, the kaon-nucleon sigma terms are key parameters for the
determination of a possible kaon condensate in dense nuclear matter \cite{Kaplan:1986yq}.

Assuming exact isospin symmetry with $m_u=m_d\equiv m_q$,  the pion-nucleon sigma term reads
\begin{eqnarray}
\sigma_{\pi N} = m_q\,\langle N(p)\,| \bar q\, q | N(p) \rangle = m_q\,\frac{\partial}{\partial m_q} m_N\,.
\label{def-sigmapiN}
\end{eqnarray}
The pion-sigma terms of the remaining baryon states are defined analogously to  (\ref{def-sigmapiN}).
As indicated in  (\ref{def-sigmapiN}), the matrix elements of the scalar quark operator are accessible via the
derivative of the nucleon mass with respect to the light quark mass $m_q$. This follows directly from the Feynman-Hellman theorem.
The strangeness sigma term $\sigma_{sN}$ of the nucleon is determined by the derivative of
the nucleon mass with respect to the strange quark mass $m_s$:
\begin{eqnarray}
\sigma_{s N} = m_s\,\langle N(p)\,| \bar s\, s | N(p) \rangle = m_s\,\frac{\partial}{\partial m_s} m_N\,.
\label{def-sigmasN}
\end{eqnarray}

\begin{table}[b]
\begin{center}
\begin{tabular}{l||l|c|c|cccc}\hline
 &  \cite{Durr:2011mp} &  \cite{Horsley:2011wr} &  \cite{MartinCamalich:2010fp} & Fit 1 & Fit 2 & Fit 3 & Fit 4\\ \hline \hline
$\sigma_{\pi N}$ & $39(4) ^{+ 18}_{- 7}  $ &$ 31 (3)(4)$ & 59(2)(17)  & 34 & 33 & 31  &31\\
$\sigma_{\pi \Lambda}$ & $29(3)  ^{+11}_{-5} $ &24(3)(4) & 39(1)(10) & 24 & 21 & 20  & 20\\
$\sigma_{\pi \Sigma}$ & $ 28(3) ^{+19}_{-3}$ & 21(3)(3)& 26(2)(5) &17 & 15 & 14 & 14\\
$\sigma_{\pi  \Xi}$ & $16(2)^{+8}_{-3}  $&16(3)(4) & 13(2)(1) & 11 & 7 & 7  & 7\\ \hline \hline
$\sigma_{s N}$ & $\;\;34(14)^{+28}_{-24}  $ &$ 71(34)(59)$ & -4(23)(25) & 43 & 24 & 2 & 3\\
$\sigma_{s \Lambda}$ & $\;\;90(13)  ^{+24}_{-38} $ &247(34)(69)  & 126(26)(35) & 238 & 194 & 191 & 194 \\
$\sigma_{s \Sigma}$ & $ 122(15) ^{+25}_{-36}$ & 336(34)(69)& 159(27)(45) & 311 & 291 & 273 & 278\\
$\sigma_{s  \Xi}$ & $156(16)^{+36}_{-38}  $&468(35)(59) & 267(31)(50) & 449  & 380 & 407 & 408 \\\hline
\end{tabular}
\caption{Pion- and strangeness sigma terms of the baryon octet states in units of MeV. }
\label{tab:sigmaterms_octet}
\end{center}
\end{table}

In Tab. \ref{tab:sigmaterms_octet} we present our predictions for the pion- and strangeness sigma terms of the baryon 
octet states for various parameter set as explained in \cite{SemkePLB2012}.
They are compared with two recent lattice determinations \cite{Durr:2011mp,Horsley:2011wr}. Our values for the non-strange 
sigma terms are in reasonable agreement with the lattice results. In particular, we obtain a rather small value for the 
pion-nucleon sigma term, which is within reach of the seminal result $\sigma_{\pi N}= 45 \pm 8$ MeV of Gasser, Leutwyler 
and Sainio in  \cite{Gasser:1990ce}. The size of the pion-nucleon term can be determined from the pion-nucleon scattering data. It requires a subtle subthreshold
extrapolation of the scattering data. Despite the long history of the sigma-term physics, the precise determination is still
highly controversial (for one of the first reviews see e.g.  \cite{Reya:1974gk}). Such a result is also consistent with
the recent analysis of the QCDSF collaboration \cite{Bali:2011ks}, which suggests a value $\sigma_{\pi N} =38 \pm 12$ MeV.
Our estimate for the strangeness sigma term of the nucleon with $\sigma_{sN} \simeq 22 \pm 20 $ MeV
is compatible with the currently most precise lattice prediction $\sigma_{sN}= 12^{+23}_{-16}$ MeV in  \cite{Bali:2011ks}. 
For the strangeness sigma terms of the remaining octet states there appears to be a striking conflict amongst the values 
obtained by the BMW and QCDSF-UKQCD groups. Our values are quite compatible with the latter. 
In Tab. \ref{tab:sigmaterms_octet}  we recall also the results of a chiral extrapolation of the recent PACS-CS data by Camalich et al.
\cite{MartinCamalich:2010fp}. The analysis is based on the baryon masses truncated at N$^2$LO with phenomenologically
adjusted values for the meson-baryon coupling constants. For almost all sigma terms we find significant differences to our results.
This may reflect the significantly much less accurate reproduction of the PACS-CS data and the physical baryon masses in  \cite{MartinCamalich:2010fp}.

\section{Summary and outlook}

In this work we reviewed the results of five different lattice groups, BMW, LHPC, PACS-CS, HSC and QCDSF-UKQCD on the baryon octet and decuplet masses. 
Using the chiral Lagrangian at N$^3$LO we obtained a universal parameter set that leads to a quantitative reproduction of the baryon masses and all 
lattice simulation data taken at quite different pairs of pion and kaon masses. While the physical masses are reproduced exactly, the lattice data are 
successfully fitted with six free parameters only. The parameters may be adjusted to the pion-mass dependence of the nucleon and omega masses of the 
BMW group together with the masses of all baryon octet and decuplet states of the PACS-CS group only. In turn, the simulation data of the HSC and 
QCDSF-UKQCD groups are reproduced quite accurately. This suggests a high level of compatibility of the different lattice data sets. For pion 
masses larger than 300 MeV there appears to be an approximate linear pion-mass dependence of all octet and decuplet baryon masses.

Based on this result, we predicted the pion and strangeness sigma terms of all baryon ground states.
In particular, we obtain a quite small pion-nucleon sigma term of about $32 \pm 2$ MeV and  also for the strangeness 
sigma term of the nucleon of about $22\pm 20$ MeV. Future even more accurate lattice simulations would help to 
consolidate our parameter set and may lead to a precise determination of the additional five flavor symmetric counter 
terms that were 
not considered yet. The latter play a decisive role in meson-baryon scattering processes at next-to-leading order already.

\end{document}